\documentclass[12pt]{article}
\def\e{{\rm e}}
 
\title{\bf 
Kertesz 
on Fat Graphs?
}
\author{ {\it W. Janke}\\
Institut f\"ur Theoretische Physik\\
Universit\"at Leipzig\\
Augustusplatz 10/11 \\
D-04109 Leipzig, Germany \\
\\
{\bf and}\\
\\
{\it D.A. Johnston} and {\it M. Stathakopoulos}\\
         Dept. of Mathematics\\
         Heriot-Watt University\\
         Riccarton\\
         Edinburgh, EH14 4AS, Scotland
         }
\begin{document}
  \maketitle
                      {\Large
                      \begin{abstract}
%
The identification of phase transition points, $\beta_c$, with the percolation 
thresholds of suitably
defined clusters of spins has proved immensely fruitful in many areas
of statistical mechanics. Some time ago
Kertesz suggested that such percolation
thresholds for models defined {\it in field} might also have measurable
physical consequences  for regions of the phase diagram below $\beta_c$,
giving rise to a ``Kertesz line'' running between $\beta_c$ and
the bond percolation threshold, $\beta_p$, in the $M, \beta$ plane. 

Although no thermodynamic singularities
were associated with this line it could still be divined by looking 
for a change in
the behaviour of high-field series for quantities such as
the free energy or magnetisation. Adler and Stauffer did precisely this
with some pre-existing series
for the regular square lattice and simple cubic lattice Ising models
and did, indeed, find evidence for such a change in
high-field series around $\beta_p$. Since there is a general dearth of 
high-field series there has been no other work along these lines.

In this paper we use the solution of the Ising model in field
on planar random graphs by Boulatov and Kazakov to carry out a similar exercise
for the Ising model on random graphs (i.e.\ coupled to 2D quantum gravity).
We generate a high-field series for the Ising model on $\Phi^4$ random
graphs and examine its behaviour for evidence of a Kertesz line.
%
                        \end{abstract} }
%
  \thispagestyle{empty}
%
%
  \newpage
%
                  \pagenumbering{arabic}

\section{Introduction}

The question of how to give a geometrical, percolation-like interpretation
of the thermal transition in the Ising model was finally resolved by 
Coniglio and Klein \cite{CK}, although the idea is already implicit in the
work of Fortuin and Kasteleyn \cite{FK} on a correlated bond-percolation model
for the $q$-state Potts model and in many ways
is a realisation of Fisher's earlier ideas
on critical droplets \cite{Fish}. The key insight was to construct spin
clusters in which like spins were joined with probability
$p = 1 - \exp ( -2 \beta)$,
where $\beta = 1/k_B T$ is the inverse temperature.
These then percolated at the correct
critical temperature $T_c$ and gave the correct thermal exponents.
The definition of such stochastic clusters
has led directly to the development of cluster algorithms
and similar geometrical representations in many statistical mechanical models.

Since Coniglio-Klein clusters are intended to access the thermal phase 
transition of the Ising model they are not usually
formulated in the presence of
an external field, since in this case the free energy becomes analytic
and the thermal transition is washed out.
However it is clear that one can still build such clusters when 
the external field is non-zero. For a given external field value $H$
these clusters will percolate at some temperature 
$T_K(H)$, and varying $H$ then traces out the so-called Kertesz line 
shown in Fig.~1 in the $M, u$ plane, where $M$ is the magnetisation and
$u = \exp ( - 4 \beta )$. 
For $H=0$ we have the usual thermal Ising transition at $u_c$, whereas for
$H \to \infty$ we recover standard bond percolation for the 
lattice in question at $u_p$.

Although a percolative transition exists across the Kertesz line
no singularities are expected in the thermodynamic quantities
related to the Ising model, since these can only 
appear for $H=0$. Nonetheless, something physical is going on
since the percolative transition is signalled by ``wrong-sign'',
say down, spin
clusters losing their surface tension.
This begs the question of whether the
Kertesz line leaves any discernible signal in measurable quantities.
This issue was discussed by Kertesz himself \cite{KZ}, and later 
investigated numerically by Stauffer and Adler \cite{SA}. If one 
considers the number of down spin clusters $n_s(u,H)$
in a non-zero up field $H$
then standard nucleation theory leads to
\begin{equation}
\log n_s ( u, H) = s \log y - s^{\sigma} \Gamma
\label{below}
\end{equation}
below the Kertesz line,
where $y = \exp ( - 2 H)$ and $\sigma=1 - 1 /d$.
The second term arises from the surface tension $\Gamma$ of
the droplets and is a direct extrapolation of the
$H=0$ ($y=1$) result.
On the other hand above the Kertesz line
a zero surface tension expression of the form
\begin{equation} 
\log n_s ( u, H) = s \log y + s \log \lambda ( u, H)
\label{above}
\end{equation}
where $\lambda<1$ is postulated. This means that
cluster numbers 
will decay as
\begin{equation}
n_s =  \exp ( - \Gamma s^{1-1/d} ) y^s
\end{equation}
below the Kertesz
line whereas they will decay as $y^s \lambda^s$
above the line.

Physical quantities such as the magnetisation can be expressed
in terms of $n_s$,
\begin{equation}
M = 1 - 2 \sum_s  s n_s ( u, H),
\end{equation}
so we can see that they can be expressed in terms of a (large-field)
expansion in $y$ using equs.~(\ref{below}) and (\ref{above}). 
Since equ.~(\ref{above}) shows that $y$ is effectively replaced
by $y \lambda$ above the Kertesz line, one might expect
the series for $M$ (and others) to converge up to some $y>1$ (since
$\lambda<1$). 

\section{Square Lattice Reprise}

In order to investigate the possible consequences of a Kertesz line
Adler and Stauffer \cite{SA}
analysed the radius of convergence of some long-standing
high-field series for the free energy and the magnetisation 
of the Ising model using
Pad\'e approximants \cite{Sykes}. 
We briefly review their results here for comparison
with the random graphs in the next section.
Taking the series for the free energy $F$ or its derivative,
we have
\begin{eqnarray}
F &=& \sum_s L_s (u) y^s, \nonumber \\
\frac{dF}{dy} &=& \sum_s s L_s (u) y^{s-1},
\end{eqnarray}
where the so-called high-field polynomials $L_s(u)$ were calculated 
up to order 15 for the square lattice in \cite{Sykes}.

The series for $F$, $\frac{dF}{dy}$ 
and the magnetisation $M$ were subjected to a
standard unbiased Dlog-Pad\'e analysis in order to estimate the radius
of convergence. If we assume a singularity of the form 
$F \sim (y - y_c)^{- \lambda}$, then
\begin{equation}
\frac{d}{dy} \log F = - {\lambda \over y - y_c}
\end{equation}
and an $[L / M]$ Pad\'e approximant to $\frac{d}{dy} \log F$
(or $\frac{dF}{dy}$, or $M$),
\begin{equation}
[L / M] = {P_L ( y) \over Q_M ( y) } = { p_0 + p_1 y + \ldots p_L y^L \over
q_0 + q_1 y + \dots q_M y^M},
\end{equation}
would be expected to show the pole at $y_c$ as a zero of $Q_M ( y)$. 
For the square lattice for instance
$u_c=3-3\sqrt{2}=0.17157 \ldots$ and $u_p = 1/4$, 
so we would expect to see a change in $y_c$ at, or around, $u_p$.

We show the results of such an analysis on $\frac{dF}{dy}$
for the estimated radius 
of convergence as a function of $u$ in Fig.~2. We can see that 
there is evidence for a change in the behaviour of $y_c$ around
$u=1/4$. In the
plotted data jumps are present in 
the $[7 / 7]$ and $[6 / 7]$ approximants at $u_p$,
and the $[5 / 5]$ and $[6 / 6]$ approximants 
cease to give a real pole term for some $u \sim u_p$. 
In Fig.~2 we have plotted the $[5 / 5]$ and $[6 / 6]$ approximants 
up to the largest real pole value obtained.
As shown in
\cite{SA} the behaviour for 
other approximants and for $F$ and $M$ is similar -- in all cases
there is evidence of an increase in the estimated $y_c$ around
$u=1/4$ or the
approximant's estimates of $y_c$ become 
complex. 

One detail that should be noted is that both here
and in the random graph case investigated below there is a tendency for 
the approximants to throw up spurious cancelling pole/zero pairs,
so one must be careful to look at both the numerator and denominator
of the approximant to make sure that one is determining the ``real''
pole term. This also introduces a certain degree of subjectivity
into proceedings since the poles and zeroes are only equal to within
numerical accuracy. All the series calculations were done with {\em exact\/}
arithmetic,
with fixed (but large) precision only being resorted to in the final stage
of obtaining the roots of the Pad\'e numerators and denominators.
 
An analysis of the series for the simple cubic lattice Ising model,
available up to order 13, in \cite{SA} 
also gave very similar results, with signs
of an increase in the estimated  radius of convergence, or instability
in the approximants,  in the
vicinity of the percolation temperature for the lattice.
The usual caveats of course apply to any such discussion, since the series
involved are quite short (and of some considerable vintage), but the cumulative
evidence of the various approximants and the results on different lattices
was felt in \cite{SA} to lend support to the Kertesz scenario.

\section{Fat Random Graphs}

Given the evidence for a Kertesz line in the Ising model on the
square and cubic lattices it is tempting to look elsewhere
for the phenomenon.
In order to carry out such an analysis for other models (or lattices)
we require a high-field expansion, which are in general rather hard to come by.
In what follows we discuss obtaining such an expansion for an exact solution to
the Ising model {\it in field} on planar (fat) random graphs which was derived
by Boulatov and Kazakov \cite{kaz,BK}. They considered the partition function
for the Ising model on a single planar graph with $n$ vertices 
\begin{equation}
Z_{{\rm single}}(G^n,\beta,H) =
\sum_{\{\sigma\}} \exp \left({\beta}\sum_{\langle i,j \rangle} G^n_{ij}\sigma_i
\sigma_j + H \sum_i \sigma_i\right)\, ,
\end{equation} 
then summed it over some suitable class $\{G^n\}$ of $n$ vertex graphs
(e.g. $\Phi^3$ or $\Phi^4$ random graphs) resulting in
\begin{equation}
Z_n = \sum_{\{G^n\}} Z_{{\rm single}}(G^n,\beta,H)\, ,
\end{equation} 
before finally forming the grand-canonical sum over differently
sized graphs
\begin{equation}
{\cal Z} = \sum_{n=1}^{\infty} \left( - 4 g c \over ( 1 - c^2 )^2 \right)^n Z_n,
\label{grand}
\end{equation}
where $c = u^{1/2} = \exp ( - 2 \beta)$.   
This last expression
could be calculated exactly as matrix integral
over $N \times N$ Hermitian matrices,
\begin{equation}
{\cal Z} = - \log \int {\cal D}\phi_1~{\cal D}\phi_2~ 
\exp \left( -{\rm Tr}\left[{1\over 2}(\phi_1^2+\phi_2^2)- c \phi_1\phi_2  - 
\frac{g}{4}( \e^H \phi_1^4 + \e^{-H} \phi_2^4)\right]  \right),
\label{matint}
\end{equation}
where the $N \to \infty$ limit is to be taken to pick out the planar diagrams
and we have used the potential appropriate for $\Phi^4$ (4-regular) random 
graphs.

When the integral is carried out the solution is given by
\begin{equation}
{\cal Z} = {1\over 2}\log {z \over g}-{1\over g}\int_0^z~{dt\over t}g(t)
+{1\over 2g^2}\int_0^z{dt\over t}g(t)^2,
\label{fullpart}
\end{equation}
where the function $g(z)$ is 
\begin{equation}
g(z)=3 c^2 z^3 +  z
\left[ \frac{1}{(1-3 z)^2} - c^2 +\frac{3z(y^{1/2}+y^{-1/2} - 2)}{(1-9 z^2)^2}
\right].
\label{geq}
\end{equation}
As we can see this is a solution in field for the Ising model on random graphs,
since $g(z)$ has been obtained in full generality for $y = \exp ( - 2 H) \ne 1$.

In order to generate a high-field series from equ.~(\ref{fullpart})
we revert the series for $g(z)$ to get a series $z(g)$. This can then be
used in equ.~(\ref{fullpart}) in order to obtain an expansion for the
partition function in powers on $g$ (i.e.\ in the number of vertices).
As we noted in \cite{JJS} it is only necessary to consider
the first of the terms in ${\cal Z} = {\cal Z}_1 + {\cal Z}_2 + {\cal Z}_3$ 
with
\begin{eqnarray}
{\cal Z}_1 & = &  {1\over 2}\log {z\over g},  \nonumber \\
{\cal Z}_2 & = & -{1\over g}\int_0^z~{dt\over t}g(t),  \\
{\cal Z}_3 & = &  {1\over 2g^2}\int_0^z{dt\over t}g(t)^2, \nonumber
\end{eqnarray}
since ${\cal Z}_k = \sum_{n} a^n_k A_n(u) g^n$ where $A_n(u)$ is 
{\it identical} for all the ${\cal Z}_k$. 

This can be traced back to the generic expression for the
partition function of any Hermitian matrix model, which is
of the form
\begin{equation}
{\cal Z} = \int_0^1 d \xi ( 1 - \xi ) \log ( f ( \xi )) + \ldots,
\label{pre}
\end{equation}
where in the case of the Ising model on $\Phi^4$ graphs
$f$ is the solution to
\begin{equation} 
g \xi = \left( \frac{ 2 g f } { c} \right) \left( { 1 \over ( 1 - \frac{ 6 g f }{ c} )^2 } - c^2  + \frac{ 6 g f } { c} { ( y^{1/2} + y^{-1/2} - 2 )
\over ( 1 - 9 \left( \frac{ 2 g f } { c} \right)^2 )^2 } \right)
+ 3 c^2 \left(\frac{ 2 g f } { c} \right)^3.
\end{equation}
The expression in equ.~(\ref{fullpart}) 
is obtained from this  by defining $z= 2 g f / c$
and integrating by parts. One finds that the coefficient
of $g^n$ in ${\cal Z}_1$ should be $(n+1)(n+2)$ times the 
full value obtained from expanding  ${\cal Z}_1 + {\cal Z}_2 + {\cal Z}_3$
from these considerations, which can be confirmed by comparison
with the results presented for low orders in \cite{stau,jan}.
For example, the coefficient of $\tilde g^2$ in \cite{stau} is found 
to be\footnote{Where $g = { ( 1 - c^2)^2 \over c } \tilde g.$}
\begin{equation}
(1 / 8 ) c^{-2} y^{-1} [ 9 + (16 c^2 + 2 c^4) y + 9 y^2 ],
\end{equation}
which is  
\begin{equation}
12 \times {\cal Z}_1 = 
12 \times (3 / 2 ) c^{-2} y^{-1} [ 9 + (16 c^2 + 2 c^4) y + 9 y^2 ]
\end{equation}
as expected.
The high-field polynomials come from an expansion of the free energy 
$F$ rather
than the partition function $\cal Z$, 
so there is one further step.  

We choose to consider
$\frac{dF}{dy}$ for calculational convenience
rather than directly taking the logarithm
to obtain $F$,
since the operation of taking a logarithm
proved to be rather memory intensive.
Since the free energy per site is defined as (absorbing a factor of $-\beta$)
\begin{equation}
F = \lim_{n \to \infty} \frac{1}{n} \log {Z_n},
\end{equation}
we can approximate this by
\begin{equation}
\frac{d}{dy}F \sim \frac{1}{n Z_n }  {d Z_n \over dy } 
\end{equation}
for some sufficiently large $n$ (in our case we take $n=32$).
The behaviour observed for $\frac{dF}{dy}$ in \cite{SA}
was in any case similar to that of $F$ and $M$.

\section{The Percolation Threshold on $\Phi^4$ Graphs}

Kazakov has already calculated the threshold on $\Phi^3$ graphs,
so it is a simple matter to repeat the calculation to obtain the result
he stated ($p_{cr} = 2/3$), but did not derive, 
for $\Phi^4$ graphs in \cite{kaz}.  We have chosen to work with $\Phi^4$
graphs rather than $\Phi^3$ graphs here because the expression for
$g(z)$ turns out to be rather more inconvenient with $\Phi^3$ graphs
and reverting this in the manner of the previous section to get ${\cal Z}_1$
is much more cumbersome than it is for $\Phi^4$ graphs.

The idea is to consider
the percolation problem as the $q \to 1$ limit of a 
$q$-state Potts
model, 
where the Potts partition function is
\begin{equation}
Z_{\rm Potts} =
 \sum_{ \{ \sigma \} } \exp ( - \tilde{\beta} {\cal H} ).
\end{equation}
with ${\cal H} =  -\sum_{\langle ij \rangle} ( \delta_{\sigma_i, \sigma_j} -1)$,
and the spins $\sigma_i$ take on $q$ values. This Potts partition
function, just as in the Ising case, can be expressed as the 
matrix integral over $N \times N$ Hermitian matrices $\phi_i$
\begin{equation}
F = { 1 \over N^2} \log \int \prod_i  {\cal D}  \phi_i \exp ( - S)
\end{equation}
where
\begin{equation}
S = { 1 \over 2 } \sum_{i=1}^{q} \phi_i^2 - \tilde c \sum_{i<j} \phi_i \phi_j 
- {g \over 4} \sum_{i=1}^q \phi_i^4.
\label{qstate}
\end{equation}
and we have used $\tilde c = 1 / ( \exp ( \tilde \beta ) + q - 2 )$
for the Potts coupling and temperature
to distinguish them from the Ising case. 
The apparent factor of two difference between the $q=2$
version of $\tilde c$ and the Ising coupling in the previous section
and the consequent factor of two in the temperature scales
is accounted for by the use of a $\sigma_i \sigma_j$ interaction
in the Ising model
and a $\delta_{\sigma_i, \sigma_j}$ interaction in the Potts model.

This in turn may be recast
into a matrix external field integral by introducing a further matrix integration
in the dummy $N \times N$ Hermitian matrix $X$,
\begin{eqnarray}
F &=& { 1 \over N^2} \log \int {\cal D} X \exp ( - { 1 \over 2} X^2 ) \\ 
&\times& \left[ \int {\cal D} \phi \exp \left(  h X \phi - { 1 +h^2 \over 2} \phi^2  
+ { g \over 4} \phi^4 \right) \right]^{q}
\label{bigpotts}
\end{eqnarray}
where $h^2= \tilde c$. In the limit $q \to 1$, the percolative probability 
$p = 1 - \exp ( - \tilde \beta)$ is related to these
parameters by $h^2= \tilde c = 1/p -1$.
The mean number of
percolative clusters per unit volume $f(p)$ may be calculated
from the quantity $\zeta(g,p)$, which 
is given in terms of the Potts free energy by
\begin{equation}
\lim_{q \rightarrow 1} {\partial F \over \partial q}  = \zeta(g,p).
\end{equation}

Evaluating $\zeta(g,p)$ in the saddle-point approximation gives
\begin{equation}
\zeta(g,p)  = {1 \over 2 N} \sum_{i=1}^N (x_i^*)^2 - { 2 \over N^2} 
\log \Delta ( x^*) - F^0(g), 
\end{equation}
where the $x_i^*$ are the saddle point values of the $X$ eigenvalues and $F^0(g)$ is the 
standard one-matrix model free energy.
As in the $\Phi^3$ model the ${ 2 \over N^2} \log \Delta ( x^*)$ term is what counts for the percolative
critical behaviour, and this gives
\begin{eqnarray}   
{ 2 \over N^2} \log \Delta ( x^*)  &=& \int_{-a}^{a}  \int_{-a}^{a} \rho (u ) \rho( v) du dv \left[
\log \left|  \left( {1 \over p } - {1 \over  2 } \right) - {1 \over  2 }
 ( u^2 + u v + v^2 )  \right|  + \log  \left| u - v \right| \right]  \nonumber \\
&-& {1 \over  2 } \log ( {1 \over p } - 1 ),
\end{eqnarray} 
where $\rho$ is the eigenvalue density for the one-matrix $\Phi^4$ model
which has support on $ [ -a, a ] $.
The $\Phi^4$ model will be critical when the argument
of the first logarithm is zero within the region of integration, which  
first occurs when $u^2  = uv = v^2 = a^2 = 2/3$, so  $p_{cr} = 2/3$ as
announced in \cite{kaz}. 
Since $p_{cr} = 2/3 = 1 - \exp ( -\tilde \beta_p)$, allowing
for the factor of two between Ising and Potts
conventions gives $\exp ( -\tilde \beta_p) = \exp ( - 2 \beta_p) = 1/3$,
so on the $\Phi^4$ graphs $u_p =  \exp ( - 4 \beta_p) = 1/9$
to be compared with the $1/4$ of the regular square lattice.

\section{Pad\'e Approximants for $\Phi^4$ Graphs}

The procedure is identical to the square lattice investigation: here we take
$Z_{32}$, calculate
\begin{equation}
\frac{d}{dy}F \sim \frac{1}{32 \; Z_{32} }  {d Z_{32} \over dy } 
\end{equation}
as our estimate of $\frac{d}{dy}F$
and then take unbiased Dlog-Pad\'e approximants to this expression
in order to  determine $y_c$. Again, exact arithmetic is used
until the final stage of extracting the zeroes and poles of the Pad\'e
approximant is reached. 

Since our starting series is rather longer than the square lattice
case, we calculate up to the $[15 / 15]$ approximant. In Fig.~3 we show
the $[13 / 13]$, $[14 / 14]$ and $[15 / 15]$ approximants plotted against $u$.
The near-diagonal approximants and lower approximants
on the $\Phi^4$ graphs which we also calculated all give very similar
estimates for $y_c$ and we have not shown them for clarity.

It is immediately obvious from Fig.~3 that although there is evidence
for a change in the behaviour of the high-field series for $u > u_p$,
it is rather less clear cut than for the square lattice series.
We have emphasised the fact by plotting the $[7 / 7]$
square lattice approximant over the $\Phi^4$ results to allow
a direct comparison. 
Indeed, without the prejudice of a previous calculation of $u_p$ it
would be fair to state that although there are some signs 
of an increase in the estimates of $y_c$ in the vicinity of $u_p = 1/9$
the variation is strongly apparent only for larger $u$ values. 

Our calculation of the series for $\Phi^4$ graphs is for 
a set
which includes various degeneracies such as self-energy and ``setting-sun''
diagrams in the ensemble of graphs, and the inclusion of such graphs
has been found empirically to reduce finite-size effects \cite{jan2}
by comparison with more restricted ensembles of $\Phi^4$ graphs.
It is of course possible that {\it any} ensemble of random graphs
may be subject to greater finite-size corrections than
a square lattice
so it is conceivable that this might explain a
weaker signal for the Kertesz line than on the square lattice.
 
\section{Conclusions}

We reviewed the existing results for the existence of the Kertesz
line for the Ising model
on the square lattice, which lend support to the suggestion that
a change in the behaviour of the high-field series should be  visible
around the percolation threshold. We then made use of the Boulatov and
Kazakov solution of the Ising model on planar random graphs to
construct a high-field series for planar $\Phi^4$ graphs and subjected this
to the same Dlog-Pad\'e analysis carried out on the (shorter) square
lattice series.  We found that there are clear signs in this series
too for a change in behaviour, since
the estimated values of the radius of convergence, $y_c$, increase
as $u = \exp ( - 4 \beta )$ is increased, as seen in Fig.~3. What is less
clear is the onset of this behaviour -- there are no large jumps in the
estimated $y_c$ at $u_p$, such as occur for at least some of the approximants
to the square lattice series. It is fair to say however, that all
the plotted approximants (and others
we have not plotted) do begin to increase  around, or just above, the
calculated value of $u_p$ for the $\Phi^4$ graphs, namely  $1/9$. 

On balance therefore, the analysis of the high-field series
for the Ising model on planar $\Phi^4$ random graphs also lends
support to the existence of a Kertesz line. Since clusters of spins
behave in the same manner on ensembles of planar random graphs as
they do on the square lattice (the Wolff \cite{Wolff} 
or Swendsen-Wang \cite{SW} algorithms, for instance, still do an excellent
job of reducing critical slowing down for Ising models 
on ensembles of random graphs) this is reassuring.
It would have been interesting, but beyond the scope of the resources available
to us, to extend the order of the $\Phi^4$ high-field series to see what effect
this had on the estimates of $y_c$. It would also be an interesting 
exercise to extend the now vintage square lattice series for further
comparison, bearing in mind that the clearest signal for a jump
in $y_c$ was only  seen in the highest available approximants.

\section{Acknowledgements}

W.J. and D.J. were partially supported by
EC IHP network
``Discrete Random Geometries: From Solid State Physics to Quantum Gravity''
{\it HPRN-CT-1999-000161}.

\bigskip
%

\clearpage \newpage
\begin{figure}[t]
\vskip 15.0truecm
\includegraphics{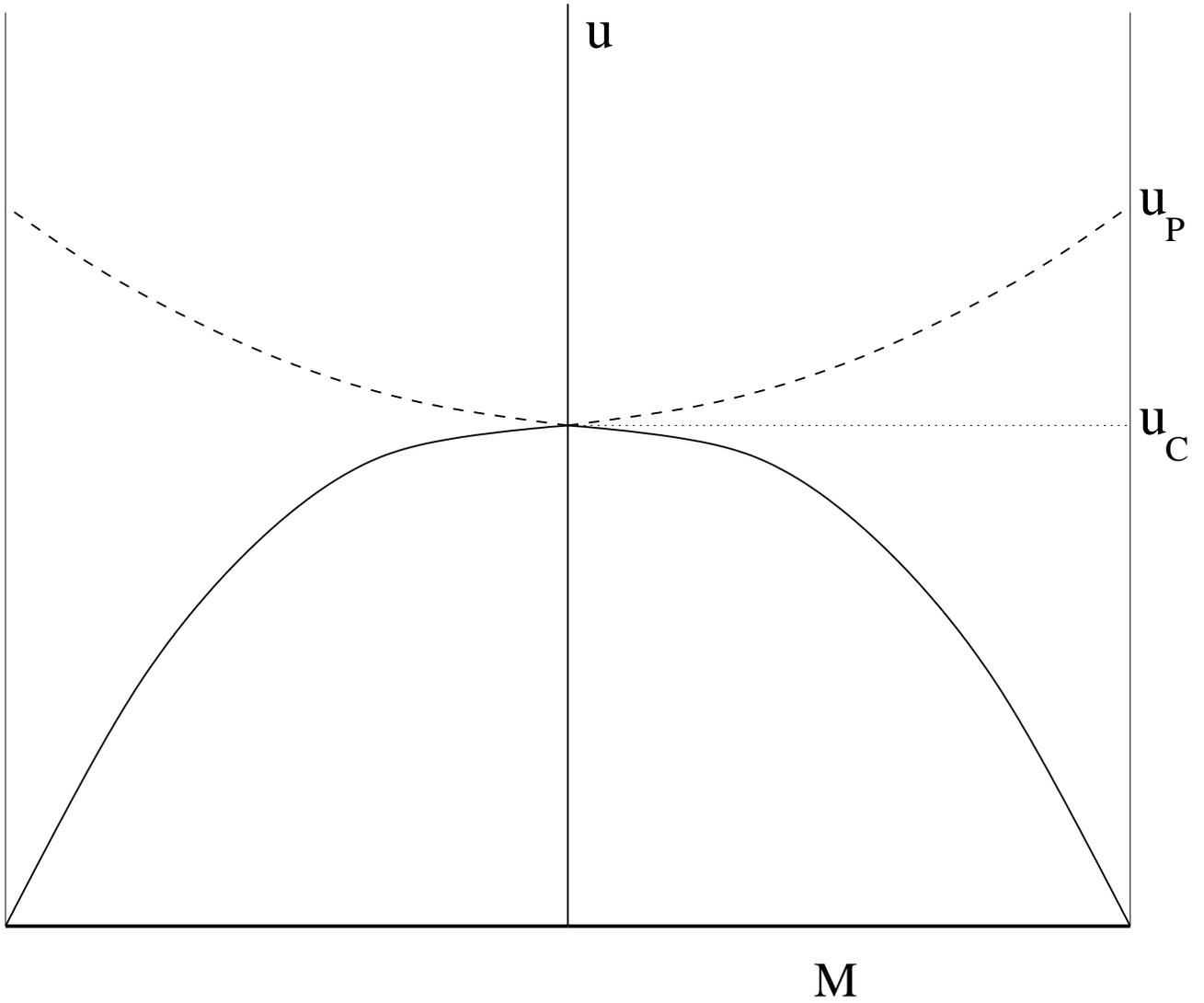}
\caption[]{\label{fig1} 
A schematic
drawing of the Kertesz line in the $M, u=\exp ( -4 \beta)$ 
plane. The phase co-existence line is shown in bold, the
Kertesz line dotted and the critical $u_c$ and percolation
$u_p$ points are marked.
}
\end{figure}  

\clearpage \newpage
\begin{figure}[t]
\vskip 14.0truecm
\includegraphics{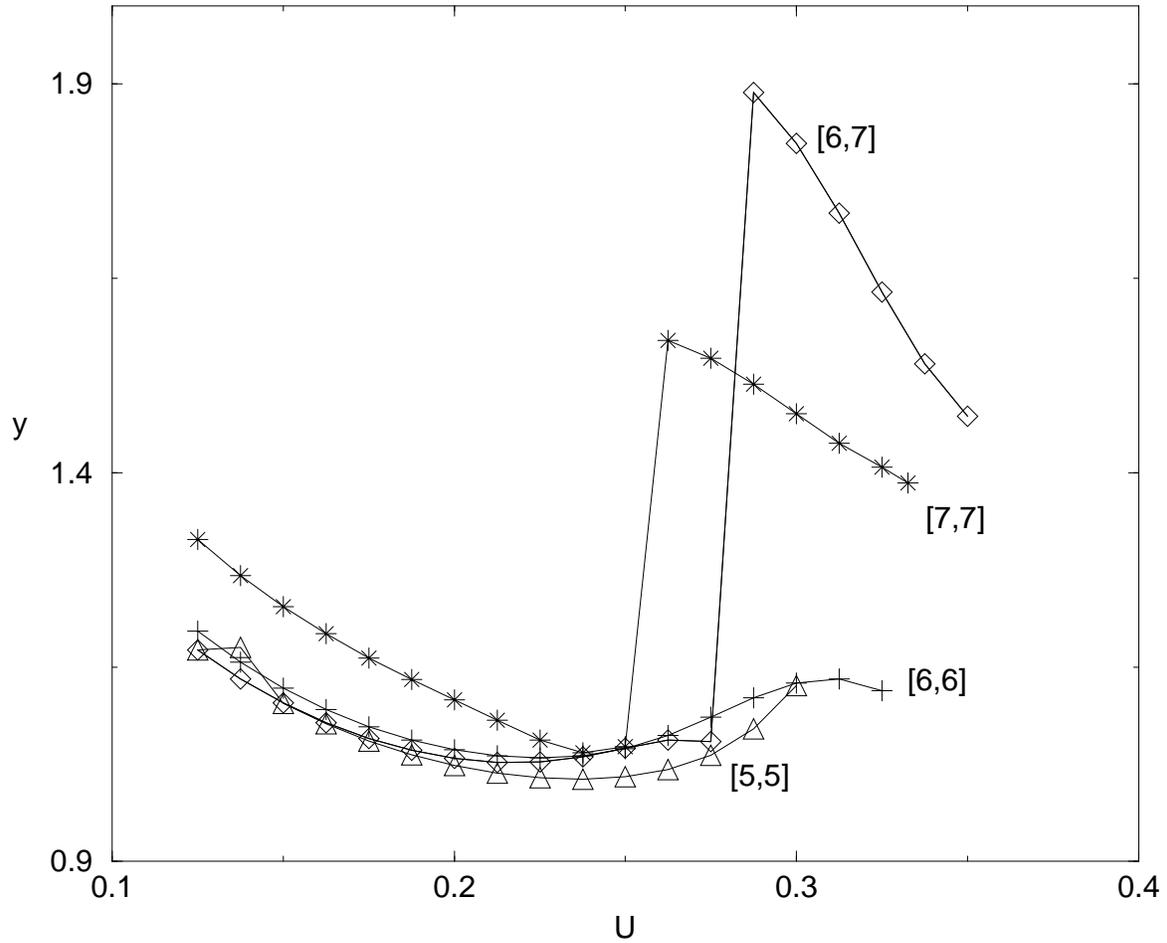}
\caption[]{\label{fig2} The estimated radius of convergence for the high-field
series of $\frac{dF}{dy}$ on the square lattice plotted against $u$. The 
different symbols
represent the different approximants used. Note that the jump is
seen in both the $[7 / 7]$ (star) and $[6 / 7]$ (diamond)
approximants, whereas the $[6 / 6]$ (cross) and $[5 / 5]$ (triangle)
approximants give
complex roots for some $u \sim u_p$.
}
\end{figure}
 

\clearpage \newpage
\begin{figure}[t]
\vskip 14.0truecm
\includegraphics{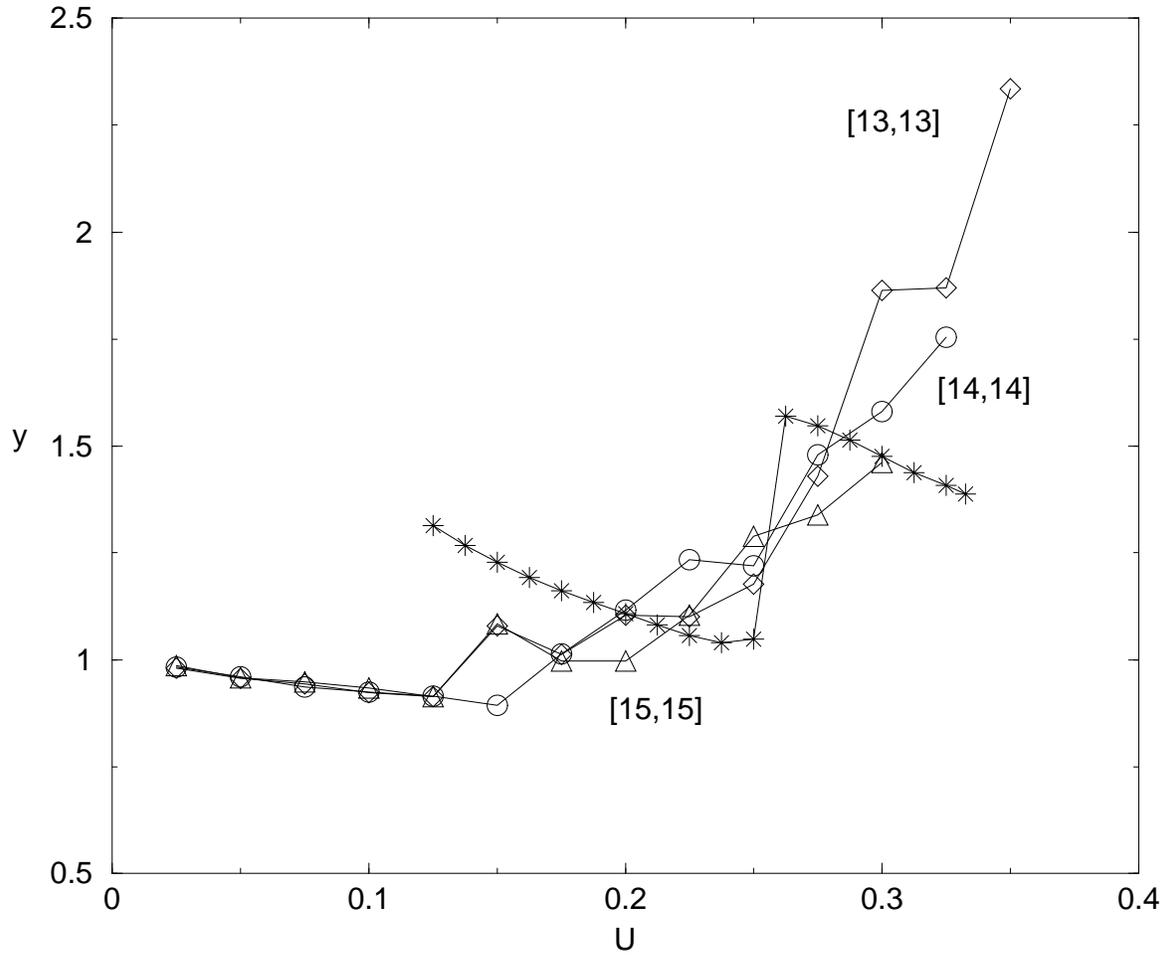}
\caption[]{\label{fig3} The estimated radius of convergence for the high-field
series of $\frac{dF}{dy}$ on $\Phi^4$ graphs plotted against $u$. The 
symbols are the estimates
obtained from $[13 / 13]$ (diamond), $[14 / 14]$ (circle) and $[15 / 15]$ 
(triangle) Pad\'e approximants, and
the $[7 / 7]$ approximant for the square lattice is also shown for
comparison using the starred symbol.
}
\end{figure}

\end{document}